\documentclass{elsart}

\usepackage{epsfig}
\usepackage{graphicx}
\def\dec{\rightarrow}
\begin{document}
\runauthor{Marina}
\begin{frontmatter}
\title{The CLEO III Ring Imaging Cherenkov Detector}
\author[Syracuse]{M. Artuso$^{1,}$}
\author[Syracuse]{R. Ayad, A. Efimov, S. Kopp, G. Majumder and R. Mountain,}
\author[Syracuse] {S. Schuh, T. Skwarnicki, S. Stone,
G. Viehhauser and J. Wang}
\author[Minnesota]{S. Anderson, Y. Kubota, and A. Smith,}
\author[smu]{T. Coan, V. Fadeyev, and J. Ye}
\address[Syracuse]{Syracuse University, Syracuse, NY 13244}
\address[Minnesota]{University of Minnesota, Minneapolis, MN 55455}
\address[smu]{Southern Methodist University, Dallas, TX 75275}
\thanks[*]{Corresponding author.\  
E-mail: artuso@phy.syr.edu}
\begin{abstract}
The CLEO detector has been upgraded to include a state of the art 
particle 
identification system, based on the Ring Imaging Cherenkov detector 
(RICH) 
technology, in order to take data at the upgraded CESR electron 
positron 
collider. The expected performance is reviewed as well as the 
preliminary
results 
from an engineering run during the first few months of
operation of the CLEO III detector. 
\end{abstract}
\
\end{frontmatter}

\section{Introduction}
A crucial element of all the experiments exploring heavy flavour 
phenomenology is the particle identification system, with the primary 
goal
of identifying $\pi$'s, $K$'s and $p$'s. CLEO III is instrumented with 
a 
Ring Imaging Cherenkov 
detector (RICH) implemented in a barrel geometry comprising a total of 
20 cm 
radial 
space between the tracking system and the CsI electromagnetic 
calorimeter.

Our goal is to achieve a $\pi/K$ separation greater than $3\sigma$ up 
to the
maximum momentum of particles produced in $B$ meson decays at the
$\Upsilon(4S)$, $\sim 2.65$ GeV/c, 
the momentum characteristic of the very important
decays $B\dec \pi\pi/K\pi$.
The $\pi/K$ separation at this momentum is 14.3 mr. Thus the required 
angular
resolution per track is $\sigma _t\le 4.8$ mr. Our design parameters 
are a
resolution per photon, $\sigma _{\gamma}$, of 14 mr and approximately
12 detected photons. This corresponds to 
$\sigma _t=4.2$ mr. Note that we
expect $\sim 2 \sigma\ \pi /K$ separation from dE/dx information from 
the 
drift chamber enclosed in our system. Our Monte Carlo simulation, 
confirmed
by an extensive set of test beam data \cite{rich-nim}, predicts that 
the CLEO
RICH detector meets our goals. These expectations will be compared
with
preliminary data from an engineering run that took place during the 
first few
months of detector operation.
\section{Detector description}
 Figure 1 shows a $r-\phi$ section of the CLEO III RICH detector 
system. It
involves an inner cylindrical shell comprising thin LiF crystal
radiators \cite{ray}, where charged particles moving with a speed
greater than the speed of light in this medium produce the Cherenkov 
cone.
Note that when the charged particle crosses the detector at normal 
incidence,
the Cherenkov light is trapped inside the radiator because of total 
internal 
reflection. We have
demonstrated \cite{sawtooth} that this problem can be circumvented by 
shaping
the outer surface of the radiator like the teeth of a saw, hence the 
name
``sawtooth radiator''. The production of these radiators is more 
difficult
because of the necessity of unconventional polishing techniques. Thus 
we are
employing them only up to a dip angle of 22$^{\circ}$.

The outer cylindrical shell of the detector is composed by thin gap 
multiwire chambers with their cathode plane finely
segmented into 7.5$\times$8 mm$^2$ pads, to  reconstruct
the location of the Cherenkov photons at the detector surface
 with high precision. The photosensitive element in these chambers is
triethylamine (TEA), dispersed in ${\rm CH}_4$. 
The sensitive bandwidth of ${\rm CH}_4$-TEA is centered
around 150 nm. Therefore we need to detect vacuum ultraviolet photons. 
Thus
the interior faces of these chambers are made of 
${\rm CaF_2}$ windows, coated with metal strips that provide one of 
the two
cathode planes. The use of ${\rm CaF_2}$ windows and the need to seal 
tightly
this system
posed a great challenge to the mechanical design. 

The cathode pad signals are processed by low
noise front end electronics devices located at the 
back of the cathode planes.  The fine segmentation needed involves a 
relatively
complex readout system, 
featuring digitization and sparsification of 230,400 charge signals. A 
low 
noise ASIC 
called VA\_RICH, custom made for this application by IDE AS, Norway,
converts the charge induced in the cathode pad into a differential
current signal. It features
excellent noise performance, measured to be :
\begin{equation}
\label{enc}
ENC(e^-)=130 +9\times C_{in}({\rm pF})
\end{equation}
This corresponds to an equivalent noise charge expected to be of the 
order of
250 $e^-$ for the input capacitance seen by the preamplifier, 
dominated by
the relatively long traces connecting the chamber pad to the
chip input. The goal of keeping the electronic noise to a minimum is 
dictated
by
the single photon response, exponential in shape for the moderate 
gains at
which we are planning to operate our chambers.
This chip features also a high dynamic range, of the order of a few 
100,000 electrons, in order to measure also the pulse height produced 
by the
minimum ionizing particle, 20-50 times higher than the single photon 
pulse.
The differential current signal is digitized and sparsifed in the data
 boards
residing in VME crates located a few meters away from the detector.
The challenge of maintaining the low noise performance in such a complex
system is mitigated by the implementation of an online common mode
subtraction prior to the sparsification process.

\section{Preliminary performance characterization}
The RICH detector was installed in CLEO in early August 1999 and the
 first data
were collected starting in November 1999. The detector performance is 
very
satisfactory. The transparency of the
expansion volume is above 99\% at 150 nm. The high voltage system is 
very stable and has been 
successfully
operated at several different voltage settings. The
average pad gain in the data reported here is $\sim 2.5\times 10^4 e^-$.
 The electronics noise, prior
 to any
working point optimization, has a mean value of 500 $e^-$, upon common
 mode
subtraction, and of 750 $e^-$ including the coherent component. 
This
performance is quite remarkable for such a large system, powered by
boards with a mixed digital and analog environment.
 
During the engineering run, the detector has been 
characterized in a
``stand alone'' mode, without resorting to any kind of external tracking
information, not yet available. The data set used for this preliminary
study is composed primarily of Bhabha
events. For these events, having the simple topology of back to back 
charged
tracks of opposite charge, we can reconstruct the track helix parameters
using the centroid of the clusters corresponding to the $e^+$ and $e^-$
intersection with the RICH pad chambers, combined with constraints 
determined
by the event kinematics and the beam properties. We selected the large 
pulse
heights to be charged track clusters. In addition we set their 
momentum equal
to the beam energy and we assumed that the two tracks originated from 
a point
with coordinates $x_0=y_0=0$, exploiting the small transverse 
beam size, and 
$z_0= (z_{e^+}+z_{e^-})/2$, where $z$ is the coordinate along the beam
direction.

Fig.~\ref{track-res} shows the Cherenkov angle distribution for
single photons. These data are fit with a signal shape 
plus a polynomial background function. We find 
$\sigma _{\gamma}=14.7\pm 0.2$ mrad for flat radiators and $\sigma
_{\gamma}=12.2 \pm 0.4$ mrad for sawtooth radiators. These results are
preliminary because the real tracking information from 
the
combined silicon strip detector and drift chamber
was not available. Moreover the RICH
alignment is still underway. The number of detected photoelectrons is
consistent with our expectations of about 12 $\gamma$'s per
track. Thus we can be confident that, once 
the tracking information is available and the alignment procedure
is complete, we will achieve excellent particle separation at all the 
momenta
of interest.

\section{Conclusions}
The first few months of operation of the CLEO III RICH have been quite
successful. The detector has performed in a reliable manner and has
proven to meet our specification. It will be a key
element in the CLEO physics program in the next few years. 

\section{Acknowledgements}
 We would also 
like to thank T. Ypsilantis for the inspiring role that he had at the
beginning of 
this work and for
the great creative energy that he continues to bring to the field of
particle identification. 
 
\eject
\begin{figure}[hbt]
\vspace{1cm}
\centerline {
\epsfig{figure=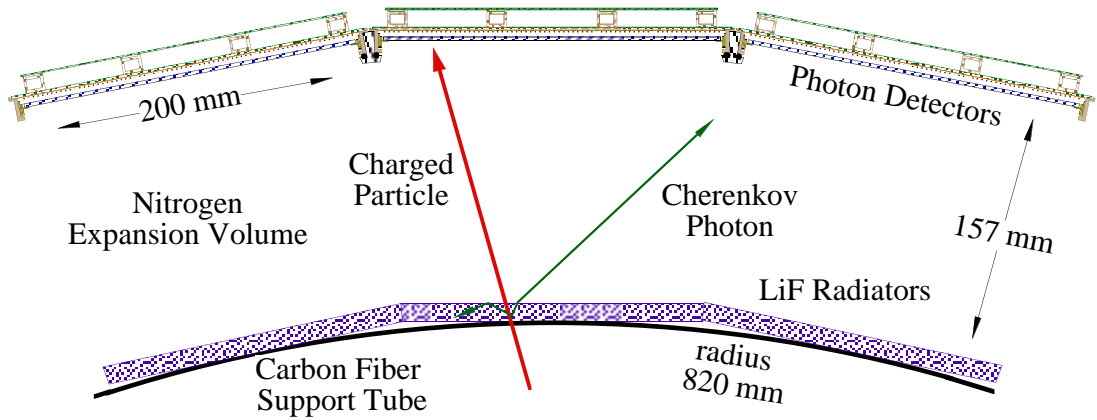,width=6in, angle=0.}}
\vspace{-4cm}
\caption{$r-\phi$ section of one tenth of CLEO RICH detector as seen
from the end.}
\label{richfig}
\end{figure}
\begin{figure}[hbt]
\centerline {
\epsfig{figure=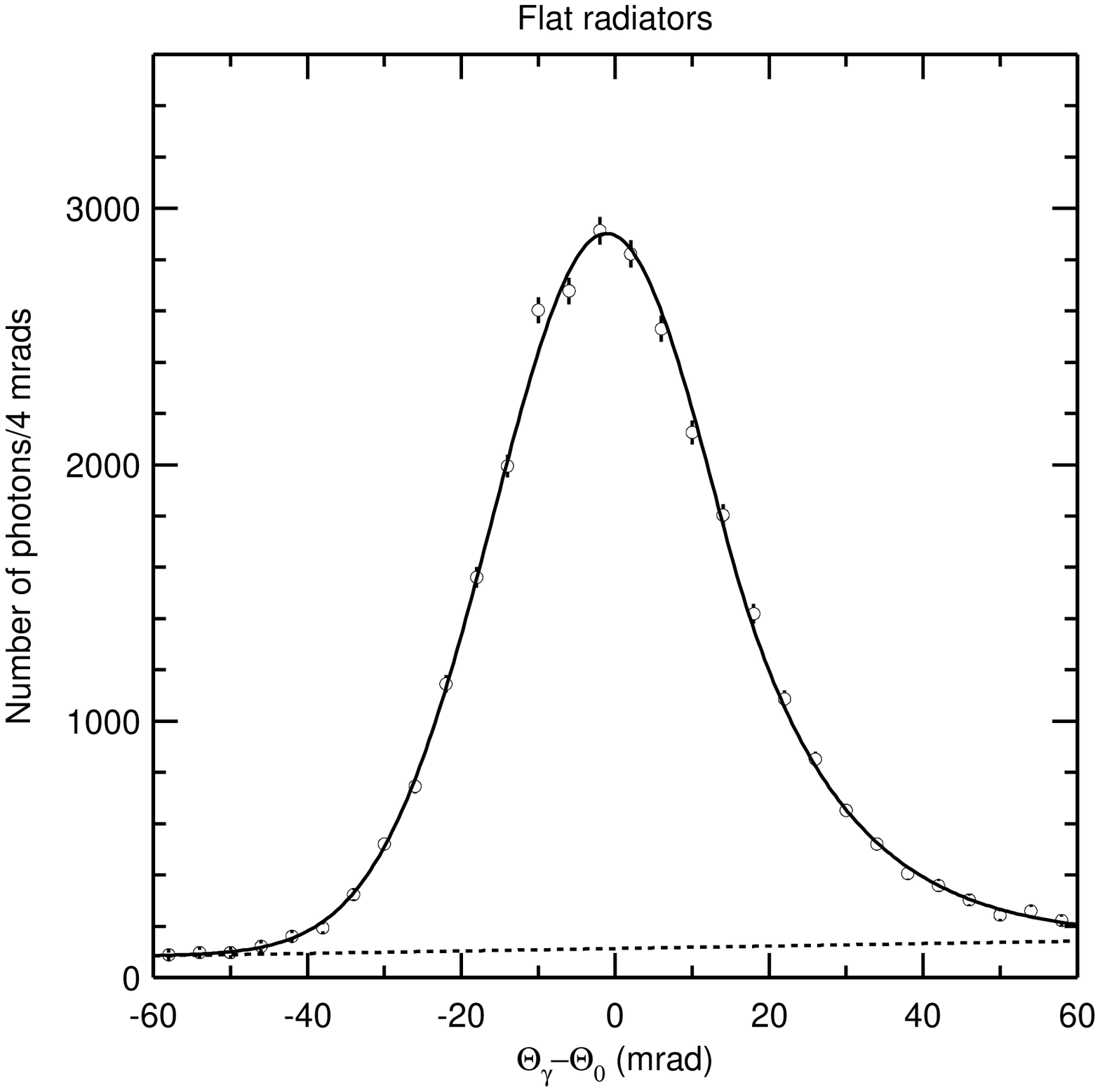,width=3in, angle=0.}
\epsfig{figure=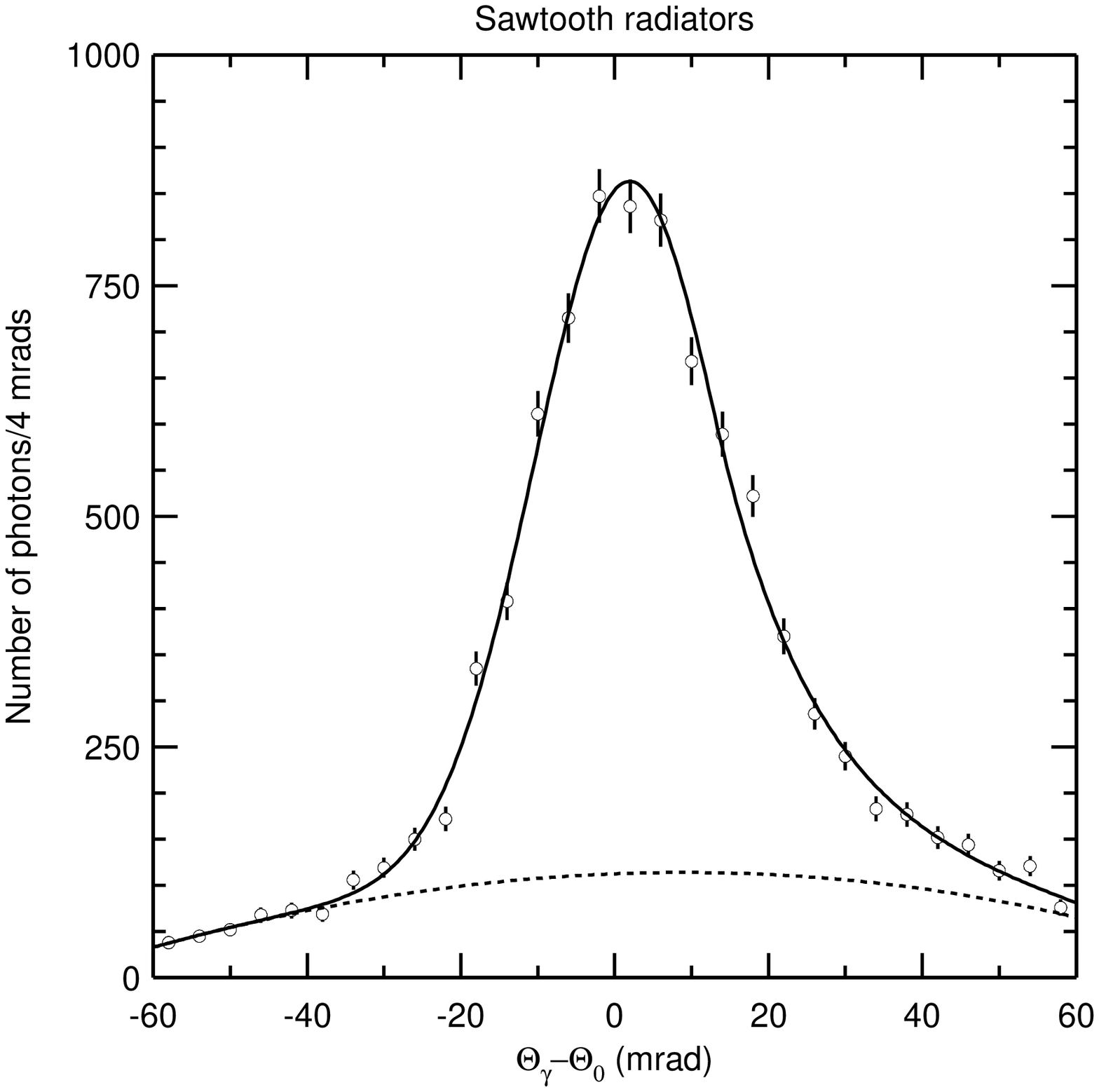,width=3in, angle=0.}}
\caption{Single photon Cherenkov angle distribution for plane radiator 
(left) and sawtooth radiator (right).}
\label{track-res}
\end{figure}

\begin{thebibliography}{9}
\bibitem{rich-nim} M. Artuso {\em et al.}, {\em Nucl. Instr. Meth. } 
{\bf
A441} (2000) 374.
\bibitem{ray} R. Mountain {\em et al.}, {\em Nucl. Instr. Meth. } 
{\bf
A433} (1999) 77.
\bibitem{sawtooth} A. Efimov {\em et al.}, {\em Nucl. Instr. Meth. } 
{\bf
A365} (1995) 285.
\end{thebibliography}
\end{document}